%% file: Top_file_journal_test.tex
\documentclass[lettersize,journal]{IEEEtran}
\RequirePackage{times}
\RequirePackage{algorithmic}
\PassOptionsToPackage{boxed}{algorithm}
\RequirePackage{algorithm}
\RequirePackage{multicol}
\usepackage{amsbsy}
\usepackage{caption}
\usepackage{helvet}
\usepackage{enumerate}
\usepackage[utf8]{inputenc}
\DeclareUnicodeCharacter{202F}{\,}
\usepackage{amsmath}
\usepackage{amsfonts}
\usepackage{graphicx}
\usepackage{multirow}
\usepackage{caption}
\usepackage{comment}
\usepackage{hyperref}
\usepackage{xcolor}
\usepackage{amsmath,amsfonts}
\usepackage{algorithmic}
\usepackage{array}
\usepackage[caption=false,font=normalsize,labelfont=sf,textfont=sf]{subfig}
\usepackage{textcomp}
\usepackage{stfloats}
\usepackage{url}
\usepackage{verbatim}
\usepackage{graphicx}
\newcommand{\ls}[1]
    {\dimen0=\fontdimen6\the\font
     \lineskip=#1\dimen0
     \advance\lineskip.5\fontdimen5\the\font
     \advance\lineskip-\dimen0
     \lineskiplimit=.9\lineskip
     \baselineskip=\lineskip
     \advance\baselineskip\dimen0
     \normallineskip\lineskip
     \normallineskiplimit\lineskiplimit
     \normalbaselineskip\baselineskip
     \ignorespaces
    }

\newcommand{\beq}{\begin{equation}}
\newcommand{\eeq}{\end{equation}}

\def\BibTeX{{\rm B\kern-.05em{\sc i\kern-.025em b}\kern-.08em
    T\kern-.1667em\lower.7ex\hbox{E}\kern-.125emX}}
\usepackage{balance}
\begin{document}
\title{Design Considerations for Phase Modulation in Testable Photonic Systems and Co-packaged Optics}
\author{\IEEEauthorblockN{
Pratishtha Agnihotri,
Priyank Kalla,
Steve Blair}\\
}

\maketitle
\input{abstract_test.tex}

\input{introduction_test.tex}
\input{Previous_work.tex}
\input{experiment.tex}

\input{application.tex}

\input{conclusion.tex}
\bibliographystyle{IEEEtran}
\bibliography{modulation}
\end{document}

%% file: abstract_test.tex
\begin{abstract}
As silicon photonic integrated circuits (PICs) scale in complexity, testing and calibration increasingly depend on effective phase modulation mechanisms. This work compares thermally induced phase modulation and carrier-based electrical modulation in Mach–Zehnder and microring modulators. The devices are designed and evaluated in terms of extinction ratio, tuning efficiency, power consumption, and modulation bandwidth. The study identifies key trade-offs between modulation speed, energy consumption, and tuning controllability, which directly influence their suitability for test signal generation and calibration tasks. The results highlight the relative advantages and limitations of thermal and electrical approaches across different operating regimes. These findings provide practical design guidance for selecting phase modulation strategies in scalable silicon photonic systems with integrated test and calibration requirements.

\end{abstract}

%% file: introduction_test.tex
\section{introduction}
Silicon photonic integrated circuits (PICs) are increasingly deployed in high-speed interconnects, data centers, and emerging computing platforms due to their ability to deliver high bandwidth and energy-efficient communication \cite{CPO, data_FPV, ML_photonics}. As these systems scale in complexity and integration density, ensuring their correct functionality becomes increasingly challenging. Unlike electronic circuits, PICs are highly sensitive to fabrication-induced variations such as waveguide width deviations, refractive index non-uniformities, and wall surface roughness \cite{pratishtha1, pratishtha12}. These variations introduce phase errors, resonance shifts, and signal degradation, which can accumulate across cascaded components and significantly impact overall system performance and yield.

Testing and calibration of PICs remain difficult due to limited observability in the optical domain and the lack of mature, automated test methodologies \cite{grating_assisted_coupling, fully_automated}. Conventional electronic testing techniques cannot be directly applied, as optical signals must be accessed, probed, and interpreted without disrupting circuit operation. As a result, scalable and low-overhead test solutions are required to enable fault localization, signal integrity validation, and post-fabrication calibration in large-scale photonic systems. This has led to growing interest in design-for-test (DfT) strategies that embed testability features directly into the photonic circuit during the design phase.

DfT architectures for silicon photonics rely on the integration of test-access points that allow selective probing of optical signals at different locations in the circuit. In prior work, carrier-depletion-based Mach–Zehnder modulators (MZMs) and microring resonators have been used as such test-access elements to extract phase and power information \cite{pratishthaitc24}. These architectures operate in dual modes, where a fraction of the optical signal is tapped and compared with a reference using on-chip interferometric structures. While effective, the choice of modulation mechanism directly impacts the efficiency, overhead, and scalability of these test-access structures.

Carrier-based electrical modulation provides high-speed phase control but incurs significant power consumption and increased device complexity due to doping and junction formation \cite{thermal_ring_sakib}. These characteristics can limit their practicality for dense DfT insertion, where multiple test-access points are required. In contrast, thermo-optic modulation offers an alternative approach by exploiting the temperature dependence of the refractive index to achieve phase tuning using localized microheaters \cite{THERMAL_mzi}. Although slower in response, thermal tuning provides fine phase control, reduced electrical complexity, and potential advantages in power-efficient test operation.

This work investigates the role of phase modulation mechanisms within photonic DfT architectures by comparing thermally induced and carrier-based modulation in Mach–Zehnder and microring modulators. The study focuses on their suitability as test-access elements, evaluating their performance in terms of phase tuning capability, extinction ratio, power consumption, and bandwidth. By analyzing the trade-offs between speed, energy, and controllability, the work provides insights into how modulation choices influence test signal generation and calibration in PICs.

The results highlight conditions under which thermal or electrical modulation is better suited for test applications and offer practical design guidelines for integrating modulation-based test-access points in scalable silicon photonic systems. These insights contribute toward the development of efficient and test-aware photonic design methodologies.

%% file: Previous_work.tex
\section{Previous Work}
Silicon photonic modulators, particularly Mach-Zehnder and ring-based structures, have been extensively studied for their suitability in high-speed and energy-efficient optical interconnects. A comprehensive co-design of a PAM-4 silicon photonic transmitter using a Mach-Zehnder modulator (MZM), is presented in \cite{mzm_PAM4}, integrating both photonic and CMOS driver circuits to achieve 25 Gbps data transmission. The paper \cite{thermal_MZM_} presents a thermal modeling framework for silicon MZMs, emphasizing the effect of substrate undercut structures on thermal tuning efficiency. The study shows that optimizing the undercut depth significantly improves modulation performance by enhancing thermal confinement and reducing power consumption. In another work, \cite{thermally_modulated_mzm} demonstrates an ultra-compact and thermally actuated MZM using vanadium dioxide (VO$_2$) as a phase-change material. The device achieves high extinction ratios and efficient switching with a small footprint, showcasing the potential of VO$_2$-based hybrid photonic systems. These advancements underscore the growing utility of thermally modulated MZMs not only in high-speed data links but also in adaptive, test-aware photonic systems.

Silicon microring modulators (MRMs) have emerged as promising candidates for high-speed, low-power optical interconnects due to their compact footprint and efficient electro-optic performance. Carrier-depletion-based MRMs have achieved remarkable modulation speeds, such as the demonstration of 25 Gbps NRZ enabled by thermal tuning \cite{thermal_ring_25gbps}. Further pushing the speed limits, a 240 Gbps PAM-4 modulator with a reduced radius and an intrinsic RC bandwidth of 110 GHz has been demonstrated \cite{thermal_ring_sakib}. While these high-speed MRMs highlight the performance of carrier-depletion modulation, thermally tunable ring modulators have also been investigated for tuning and control purposes. For example, thermal tuning enables precise resonance alignment, which is vital for reconfigurability and calibration in photonic circuits. A recent thermal modeling study incorporating substrate undercut structures has shown improvement in heater efficiency by minimizing thermal leakage into the substrate, thereby reducing power consumption and improving tuning response \cite{THERMAL_RING_undercut}. These advancements demonstrate that both carrier-depletion and thermal tuning techniques are essential in enabling scalable, high-performance photonic integrated circuits (PICs), with each offering complementary benefits in terms of modulation speed and wavelength selectivity.

%% file: experiment.tex
\section{Design of thermally modulated DFT Circuitry}
To extend our previously proposed design-for-test (DFT) architectures for silicon photonic circuits, we design and evaluate thermally tuned Mach-Zehnder modulator (MZMs) and microring resonators. 

By leveraging the thermo-optic effect in silicon, precise and stable phase control can be achieved through localized heating. In this work, we present the design, simulation, and comparative analysis of these thermally tuned modulators, building upon our earlier work which utilized carrier-depletion-based tuning. The following subsections detail the working principles, layout configurations, and performance characteristics of both the thermally tuned MZM and ring modulator implementations.
\subsection{Design of Thermally Tuned MZM}
Silicon based MZMs are widely employed as phase modulators owing to their broadband, low-loss, and CMOS-compatible architecture. The performance of these devices critically depends on the method used for phase modulation. In prior work, we implemented and characterized an MZI modulated via the carrier depletion effect \cite{pratishthaitc24}. While this approach enables high-speed modulation, it has high power requirement.

In this study, we present a comparative analysis between carrier depletion and thermal tuning techniques for phase modulation in MZMs. Thermal tuning utilizes microheaters placed above the waveguide to locally change the refractive index via the thermo-optic effect. Although slower in response, thermal tuning uses smaller amount of power for signal modulation. Our goal is to evaluate both modulation methods in terms of phase shift characteristics, voltage requirement, footprint, and bandwidth using consistent design and simulation tools.

The MZM designs for both modulation techniques were modeled using the Ansys Lumerical tool suite \cite{Lumerical}. Both configurations were fabricated on a silicon-on-insulator (SOI) platform with a 500nm X 220nm silicon device layer and a 2$\mu$m oxide cladding.

We design MZM employing two 2 × 2 directional couplers
and a phase modulator. The design and simulation of 2 × 2 directional coupler and electrically tuned phase modulator is explained in our previous work. In this work, the phase modulator region is 300$\mu$m long. The phase of the input signal is modulated thermally. The thermal tuning of MZM is achieved by changing the temperature $T$ in one of its arms, and
hence its effective refractive index, $n_{eff}$. This, in turn, leads to a change in the phase of the optical signal according to Eq.\ref{eqn} \cite{Samland:25}.

\begin{align}
\Delta \phi =\frac {2\pi n_{eff}} {\lambda} L =
& \frac {2\pi} {\lambda} L\cdot \frac{\partial n}{\partial T} \Delta T,
 \label{eqn}
\end{align}
where $\lambda$ is the wavelength, $L$ is the heated arm length and $\Delta T$ the change in temperature experienced by the waveguide. The thermo-optic coefficient of silicon, $\frac{\partial n}{\partial T} = 1.8 X10^{-4} K^{-1}$ at 300K.

\begin{figure}[hbt]
  \centering
  \includegraphics[scale=0.4]{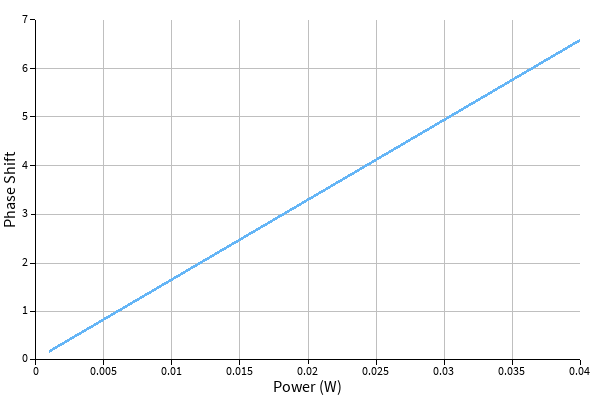}
  \caption{Phase shift in optical signal with power in the modulation arm}
  \label{mzi_thermal_power_phase}         
\end{figure}

\begin{figure}[hbt]
  \centering
  \includegraphics[scale=0.4]{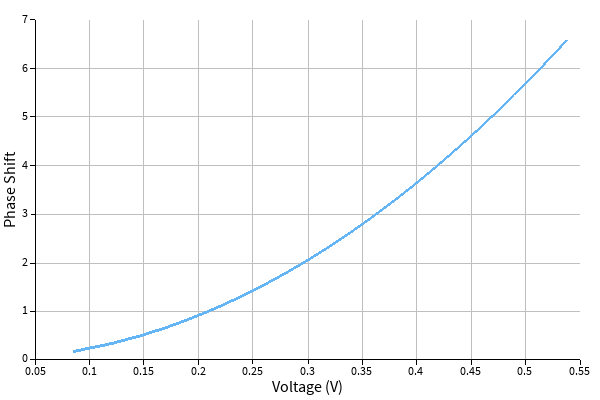}
  \caption{Phase shift in optical signal with voltage}
  \label{mzi_thermal_voltage_phase}         
\end{figure}

In our design, a rectangular aluminium wire of length 300$\mu$m, width 1.1$\mu$m, and height 1$\mu$m is used as heater. It is placed above the upper arm of the phase modulator region. The gap between the waveguide and heater is 1$\mu$m. 
A thermal simulation is performed using the HEAT solver, with power values ranging from 1mW-40mW. The temperature profile at the waveguide cross-section for different input power values is obtained. Using the Finite-Difference Eigenmode (FDE) solver, we obtain the temperature-dependent effective index of the waveguide’s fundamental mode by sweeping the input power and calculate the corresponding phase shift. A sweep over heater power is used to extract the phase shift $\Delta$$\phi$ induced in the modulator as shown in Fig.\ref{mzi_thermal_power_phase}. \\
The voltage, $V$, is calculated using $V=\sqrt{R*P}$, where $R$ is the resistance experienced by the wire and $P$ is the power consumed. The resistance, $R$, is calculated using $R= \rho \frac{L}{A}$. Here, $\rho = 2.65 X 10^{-8} \Omega$m is resistivity, $L$ is length, and $A$ is the cross-sectional area of the wire. The phase shift vs voltage behaviour is shown in Fig.\ref{mzi_thermal_voltage_phase}.
\\
The performance characteristics of the MZM are calculated. The extinction ration (ER) is defined as,
\begin{align}
 \label{er}
ER = 10log_{10}\Big(\frac{T_{max}}{T_{min}}\Big) dB,
\end{align}
where $T_{max}$, and $T_{min}$ are maximum and minimum transmission respectively. The bandwidth is calculated as $f_{max} - f_{3dB}$, where $f_{max}$ is the frequency at the maximum transmission and  $f_{3dB}$ is the frequency at which transmission falls to 0.707$T_{max}$. The $f_{max}$ and $f_{3dB}$ for the simulated MZM are 193.510THz and 193.550THz respectively. We use Non-Return-to-Zero (NRZ) format for data transmission, so maximum bit rate is twice the bandwidth.  

\begin{table}[hbt]
\begin{center}
  \caption{Characteristic parameters of modulation techniques in Mach-Zehnder modulator}
  \label{tab:MZM}
  \begin{tabular}{|c|c|c|c|}
    \hline
    \textbf{Parameter} & \textbf{Carrier Depletion} &\textbf{Thermal Tuning} \\
    \hline
    Extinction ratio &20.3dB & 29.26dB\\
    \hline
    Footprint & 0.04301mm$^2$& 0.00306mm$^2$\\
    \hline
    Modulation arm length &5mm &300$\mu$m\\
        \hline
    Voltage needed for $\pi$ phase shift &5.6V &0.39V\\
    \hline
    Maximum bit rate &80Gbps& 30Kbps\\
    \hline
    3dB-Bandwidth&40GHz& 15KHz\\
    \hline 
  \end{tabular}
\end{center}
\end{table}

In comparison to carrier-depletion-based modulator, thermally tuned MZMs offer better extinction ratio, smaller footprint, and lesser voltage requirement, making them highly effective for test, and calibration tasks within DFT architectures. While their slower modulation speed limits their use in high-speed data applications, their low design complexity and robustness to fabrication variations position them as a practical and reliable alternative for signal integrity verification and fault detection in silicon photonic circuits.

\subsection{Design of Ring Modulator}
\begin{figure}[hbt]
  \centering
  \includegraphics[scale=0.45]{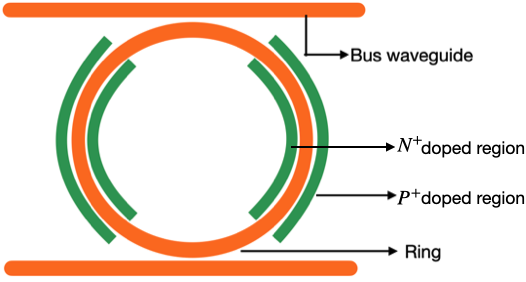}
  \caption{Ring Modulator}
  \label{ring_modulator}         
\end{figure}

In our previous work, we employed static microring resonators as fixed-wavelength filters within a design-for-test (DFT) architecture for silicon photonic circuits. While effective for fault detection at a predefined operating point, these static rings have huge power requirements for  tuning and detuning. In this work, we extend that concept by introducing actively tunable ring modulators as shown in Fig. \ref{ring_modulator}, where the resonance wavelength can be dynamically adjusted. We explore two widely used tuning mechanisms—carrier depletion and thermal tuning—to modulate the phase and shift the resonance of the rings.

\subsubsection{Modulation using Carrier Depletion Technique}

\begin{figure}[hbt]
  \centering
  \includegraphics[scale=0.25]{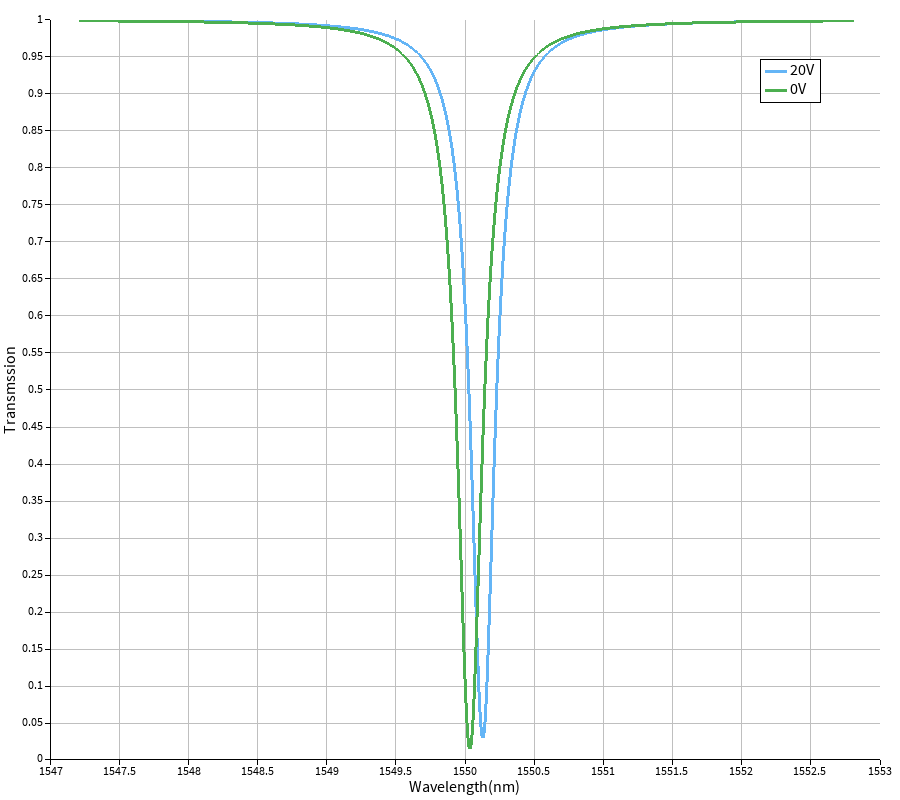}
  \caption{Shifting of resonance peak using carrier depletion modulation technique}
  \label{ring_shift_1}         
\end{figure}
We conducted detailed simulations on a silicon ring resonator of  44.42$\mu$m circumference in our previous work \cite{pratishtha_ets}. We now evaluate the performance of the carrier depletion-tuned ring modulator, ensuring continuity in device geometry. The ring resonator is designed to resonate at 1550nm, the operating wavelength. The modulation relies on changes in carrier concentration under reverse bias leading to variations in the effective refractive index, thereby shifting the resonance wavelength. 
The effective index $n_{eff}$ of the waveguide profile is obtained using FDE solver. The perturbation of the effective index with applied voltage can be characterized using the CHARGE and FDE solvers. The voltage is applied at the outer aluminium metal strip and the inner strip is connected to ground. The resonance peak shifts to the right with application of voltage as shown in Fig. \ref{ring_shift_1}. The shifting of resonant peak ensures that the ring is detuned and it no longer filters out 1550nm optical signal.
\subsubsection{Modulation using Thermal Tuning Technique}
\begin{figure}[hbt]
  \centering
  \includegraphics[scale=0.43]{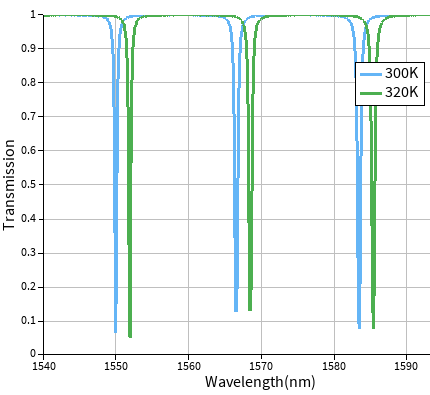}
  \caption{Shifting of resonance peak with temperature}
  \label{ring_shift}         
\end{figure}

\begin{figure}[hbt]
  \centering
  \includegraphics[scale=0.35]{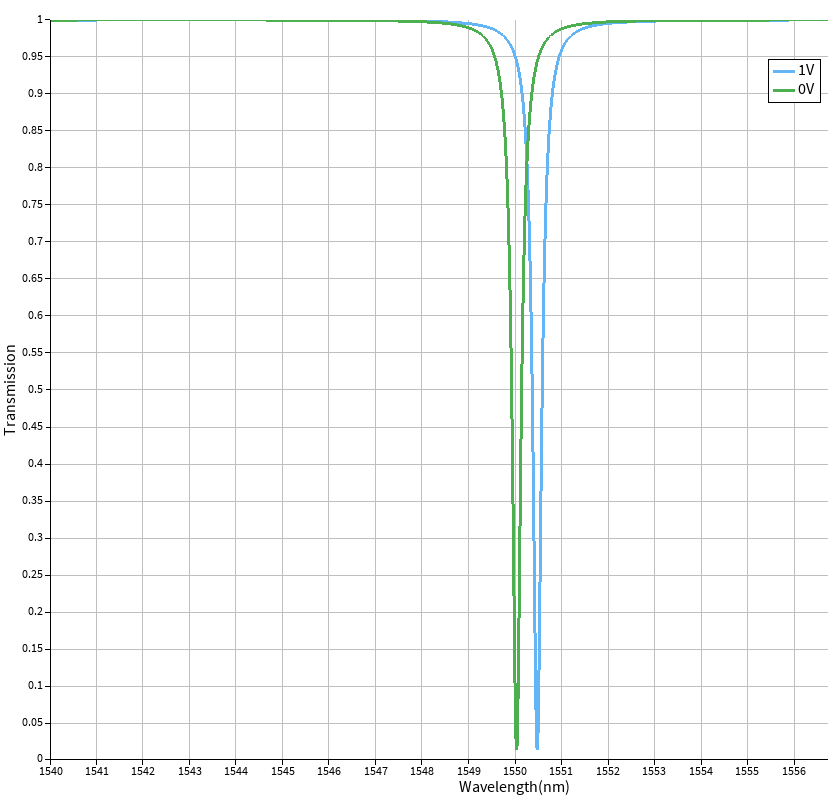}
  \caption{Shifting of resonance peak using thermal tuning modulation technique}
  \label{ring_shift_temp}         
\end{figure}

By exploiting the thermo-optic effect in silicon—where the refractive index increases with temperature—thermal tuning enables dynamic adjustment of the effective index of the waveguide. This leads to a shift in the resonant wavelength of the ring resonator, allowing for reconfiguration and compensation of fabrication-induced variations. The shifting of resonant peak in the ring modulator is observed using varFDTD. The ring is simulated at different temperatures and the observation is made as shown in Fig.\ref{ring_shift}. A thermal simulation is performed using the HEAT solver and corresponding change in refractive index vs voltage data is extracted using FDE solver. This data is further exported to INTERCONNECT. The resonance shift can be expressed by Eq. \ref{eqn_temp}. 
\begin{align}
\Delta \lambda =\frac {\lambda} {n_g} L \Big(\frac{\partial n_{eff}}{\partial T}\Big)\Delta T
 \label{eqn_temp}
\end{align}
Here, $n_{eff}$ is the effective index and $n_g$ is the group index, $\Delta \lambda $ shift in the resonant peak and $\Delta T$ the change in temperature.
The voltage vs transmission in ring modulator at different voltages is shown in Fig.\ref{ring_shift_temp}. 

\begin{table}[hbt]
\begin{center}
  \caption{Characteristic parameters of modulation techniques in ring modulator}

  \begin{tabular}{|c|c|c|c|}
    \hline
    \textbf{Parameter} & \textbf{Carrier Depletion} &\textbf{Thermal Tuning} \\
    \hline
    Extinction ratio &18.89dB & 21.32dB\\
    \hline
    FWHM& 0.21nm& 0.3nm\\
    \hline
   Modulation efficiency &4.5pm/V&40pm/V\\
    \hline
    Quality factor &7109.69& 5166.66\\
    \hline
  \end{tabular}
 \label{ring_data}
\end{center}
\end{table}

Thermally tuned modulators exhibit higher modulation efficiency, as shown in Table \ref{ring_data}. This enables substantial resonance wavelength shifts with relatively low voltage, making them suitable for applications requiring fine, static tuning/detuning. Our architecture operates in both normal and test modes, where the modulators remain detuned during standard operation and are selectively tuned into resonance to access specific blocks during testing. This mechanism allows a centralized test engine to observe and evaluate localized circuit responses with minimal impact on system performance. Simulations confirm that these thermally actuated devices achieve high modulation efficiency and low power consumption, with favorable extinction ratios and tuning ranges suitable for integration in scalable DFT frameworks.

%% file: application.tex
\section{Applicability of Thermally Tuned Modulators in DFT Architectures}
\begin{figure*}[hbt]
  \centering
  \includegraphics[scale=0.38]{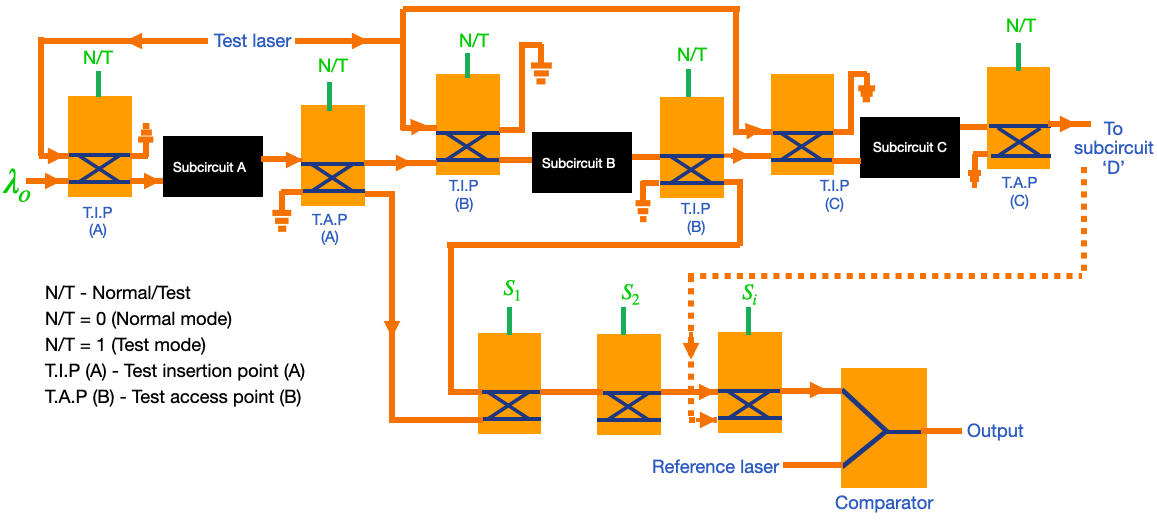}
  \caption{Thermally tuned MZM in DFT architecture}
  \label{mzi_thermal_dft}         
\end{figure*}

\begin{figure*}[hbt]
  \centering
  \includegraphics[scale=0.46]{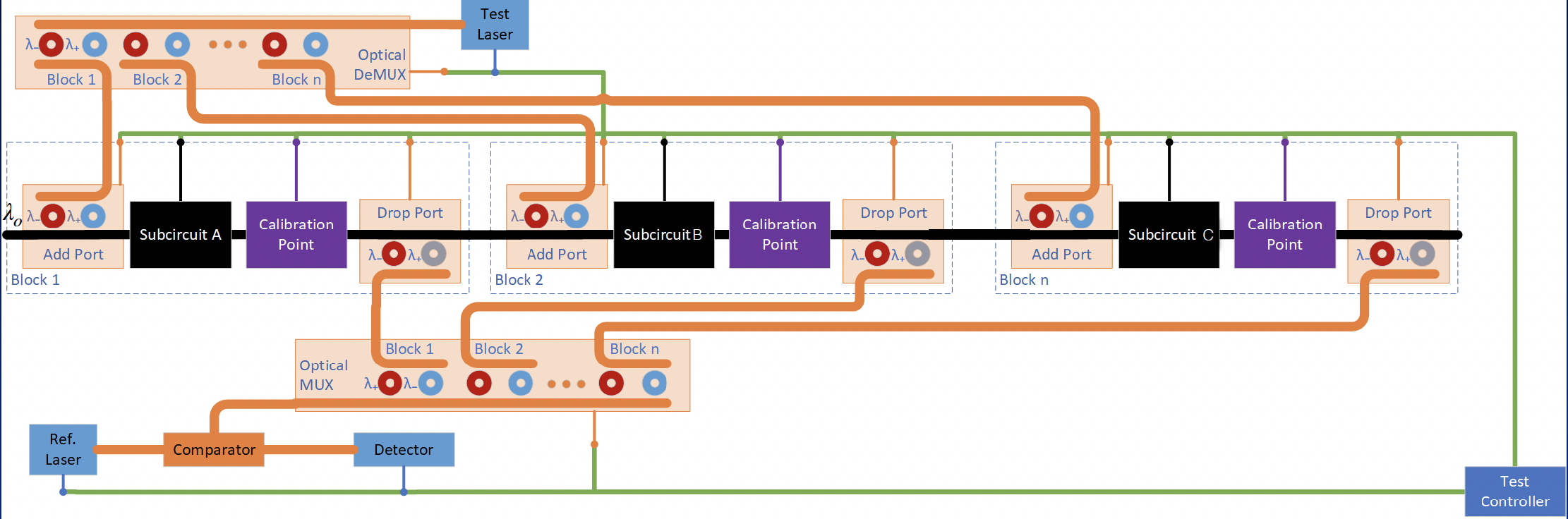}
  \caption{Thermally tuned ring modulator in DFT architecture}
  \label{ring_thermal_dft}         
\end{figure*}

In this work, we show that thermally tuned Mach-Zehnder Modulators (MZMs) and ring modulators can be effectively used in Design-for-Test (DFT) applications for silicon photonics. Thermally tuned modulators offer a simple and reliable way to shift the phase or wavelength of light by applying heat, without the need for high-speed electrical signals. MZMs are useful for modulating light over a wide range of wavelengths, while ring modulators are smaller and more selective, making them ideal for compact layouts. By using both types of modulators in a thermal tuning scheme, we can create flexible and practical test structures that help verify photonic circuits during operation.
\subsection{Thermally tuned MZM in DFT architecture}

A PIC operating at $\lambda_o$ is divided into number of subcircuits as shown in Fig. \ref{mzi_thermal_dft}. The thermally tuned MZMs are placed as test access points, test insertion points, and MUXs to function as controllable power tapping elements for DFT purposes. The MZMs can operate in two modes: normal mode and test mode. During normal operation, the MZM allows the optical signal to pass through the next connected subcircuit. In test mode, the test laser is fed into the MZM through test insertion point. The thermal tuning is applied via $N/T$ control signal to induce a phase shift, which allows a portion of the test signal power to be tapped out at test access point and routed to MUX and comparator for monitoring. The select line $S_i$ of the MUX selects the test signals from subcircuits as inputs. The power and phase of the selected test signal is compared against the corresponding parameters of the reference signal at the comparator. Here, Y-combiner is used as a comparator. If the power and phase of the test signal are similar to that of the reference signal, both signals destructively interfere. No signal at the comparator output indicates a defect-free signal, and malfunctioning otherwise. This architecture enables targeted signal extraction from specific subcircuit under test, improving test coverage while maintaining minimal impact on normal circuit operation.

\subsection{Thermally tuned ring modulator in DFT architecture}

The proposed architecture, in Fig. \ref{ring_thermal_dft}, partitions the photonic integrated circuit (PIC) into multiple subcircuits. Each subcircuit is associated with a thermally tuned ring modulators, resonating at test wavelengths, $\lambda_-$ and $\lambda_+$, that serves as a selective test access point. Under normal operating conditions, all ring modulators are thermally detuned from the operating wavelengths, allowing optical signals to propagate through the desired path without interference. During test mode, the ring modulators in optical DeMux, add port, drop port and optical MUX corresponding to the target subcircuit are thermally tuned into resonance, enabling them to couple in or tap out test signals at the specific location. This selective activation ensures that only one test path is active at a time, minimizing crosstalk and preserving circuit integrity. A control signal is used to determine which subcircuit to test by tuning the associated ring modulator, and a multiplexer (MUX) routes the captured test response to a comparator. By using thermally tuned rings instead of electrically driven ones, the architecture benefits from reduced design complexity, compact layout, and the ability to perform wavelength-selective testing without introducing significant power consumption.

Thermally tuned modulators offer a highly practical and reconfigurable solution for Design-for-Test (DFT) architectures in silicon photonics. Their ability to precisely modulate phase or shift resonance with simple thermal biasing enables selective access to internal test access points without disrupting normal operation. Despite their limited speed, their compact footprint, and low power requirements make them ideal candidates for scalable and low-overhead test and calibration strategies in complex photonic integrated circuits.

The use of heaters may change the thermal gradient on-chip. The thermal characterization abstraction model for opto-electronic layout presented has been presented by our research group \cite{thermal_ring}. The compact 2.5D thermal modeling approach enables efficient estimation of temperature gradients across photonic chips caused by electronic hot-spots. By coupling this temperature data with refractive index variation, the phase shift in modulators such as MZMs and ring resonators can be predicted. This becomes particularly useful in DFT settings, where thermal tuning is employed deliberately to modulate optical signals for test pattern generation or fault detection. The abstraction model can be adapted to simulate the impact of thermal perturbations on modulator behavior, allowing the design of test schemes that are both thermally aware and scalable.

%% file: conclusion.tex
\section{conclusion}
This work presents a DFT architecture that incorporates thermally tuned Mach-Zehnder and ring modulators to enhance testability and calibration in silicon photonic integrated circuits. By leveraging the thermo-optic effect in silicon, the proposed approach enables precise, voltage-controlled tuning of optical paths for signal tapping, isolation, and monitoring at the subcircuit level. Thermally tuned MZMs serve as broadband, low-loss phase shifters suitable for hierarchical access, while ring modulators provide compact, wavelength-selective test points ideal for dense layouts.

Incorporating thermally tuned modulators into DFT not only reduces area and design complexity compared to high-speed carrier-based schemes but also provides a robust path toward automated optical test solutions. The proposed architecture paves the way for more reliable, maintainable, and test-aware photonic systems capable of meeting the demands of large-scale commercial deployment. 

The future work will focus on developing automated test pattern generation (ATPG) algorithms tailored for thermally tunable modulators, enabling efficient detection of parametric faults in large-scale photonic integrated circuits.